\documentclass{Interspeech}

\interspeechcameraready

\title{An interpretable speech foundation model for depression detection by revealing prediction-relevant acoustic features from long speech}

\author[affiliation={1}]{Qingkun}{Deng}
\author[affiliation={2}]{Saturnino}{Luz}
\author[affiliation={3}]{Sofia}{de la Fuente Garcia}

\affiliation{Division of Psychiatry}{The University of Edinburgh}{UK}
\affiliation{Usher Institute}{The University of Edinburgh}{UK}
\affiliation{Centre for Medical Informatics, Usher Institute}{The University of Edinburgh}{UK}

\email{qdeng2@ed.ac.uk}

\usepackage{comment}

\begin{document}

\maketitle

\renewcommand{\thefootnote}{}
\footnotetext{This paper has been accepted for presentation at \textit{INTERSPEECH} 2025.}
\renewcommand{\thefootnote}{\arabic{footnote}}

\begin{abstract}
    
Speech-based depression detection tools could aid early screening. Here, we propose an interpretable speech foundation model approach to enhance the clinical applicability of such tools. We introduce a speech-level Audio Spectrogram Transformer (AST) to detect depression using long-duration speech instead of short segments, along with a novel interpretation method that reveals prediction-relevant acoustic features for clinician interpretation. Our experiments show the proposed model outperforms a segment-level AST, highlighting the impact of segment-level labelling noise and the advantage of leveraging longer speech duration for more reliable depression detection. Through interpretation, we observe our model identifies reduced loudness and F0 as relevant depression signals, aligning with documented clinical findings. This interpretability supports a responsible AI approach for speech-based depression detection, rendering such tools more clinically applicable.
\end{abstract}

\section{Introduction}

Depression is a common mental disorder, characterised by prolonged low mood, and loss of interest in activities, with an estimated 5\% of adults suffering from it globally \cite{WHO2022}. Recently, more research attention has been placed on developing automatic depression screening tools, using Deep Neural Networks (DNNs) to analyse patients’ speech. These tools have the potential to monitor individuals’ risk of depression at early stages and assist clinicians in providing rapid interventions. This paper addresses two issues that may hinder the clinical applicability of such clinical tools: predictions relying on short speech segments and a lack of model interpretability.

First, segmentation of data sequences is a common approach to avoid processing long sequences for DNN-based depression detection. For instance, \cite{sardari_audio_2022} and \cite{muzammel_end--end_2021} segmented audio data sequences along the temporal dimension, each labelled according to the subject's overall diagnosis. However, assuming that there are speech segments from depressed patients that contain \textit{no} depression-relevant information (i.e. if depressive markers are \textit{not} evenly spread across the whole speech), and are nonetheless labelled ``depressed'', this segment-level labelling approach can add noise to the model training and consequently reduce prediction accuracy in clinical practice. 

To avoid labelling noise in social media analysis, \cite{bucur_end--end_2022} applied a post-level encoder to first encode social media posts from the same user into fixed-size embeddings. The sequence of embeddings from each user is then fed to a user-level encoder for a user-level depression classification given a single label. While segment-level labelling noise may be obvious in social media analysis (i.e., text modality), where some posts do not explicitly contain depression-related utterances, its impact on speech-based depression detection (i.e., audio modality) demands further investigation.

Second, speech-based depression detection tools are hardly interpretable, hindering their clinical implementations in practice \cite{squires_deep_2023,rudin_stop_2019}. In text-based depression detection (e.g., social media analysis), attention scores from transformer models have been used for model interpretation because they provide understandable weight distributions over input features \cite{wiegreffe_attention_2019}. For instance, \cite{zogan_explainable_2022} acquired attention scores from their  Hierarchical Attention Network (HAN) to infer the importance of tokens in social media tweets for depression detection. Intuitively, input tokens at the sequence positions with high attention scores contribute more to depression detection. 

However, this interpretation approach is insufficient because it ignores the computations that happened before the attention layers \cite{chefer_generic_2021}. Specifically, in \cite{zogan_explainable_2022}, before attention operations, each token has already been contextualised by every other token by a bidirectional Gated Recurrent Unit (biGRU). Therefore, attention scores derived from later layers do not map directly onto the input tokens, but rather onto their context-enriched representations, rendering the interpretation less precise. Additionally, raw attention scores often poorly correlate with gradient-based feature significance \cite{jain_attention_2019}, questioning their validity for model interpretation. More importantly, this attention-based interpretation approach is not readily applicable to speech-based depression detection because unlike text, which is directly understandable to humans, raw audio signals require further processing to be human-interpretable.


To address these insufficiencies, we design an interpretable speech foundation model approach, leveraging pre-trained Audio Spectrogram Transformer (AST) \cite{gong_ssast_2022}, to make end-to-end depression detection using one's long speech instead of short segments. We compare the model to a segment-level AST to highlight the impact of segment-level labelling noise and the advantage of using longer speech duration for more reliable depression detection. Importantly, based on gradient-weighted attention maps \cite{chefer_generic_2021}, we introduce a novel frame-based interpretation method to extract human-interpretable acoustic features relevant to depression detection, making such speech foundation models interpretable for clinicians, and therefore more clinically applicable, for the first time.




\begin{figure*}[t]
  \centering
  \includegraphics[width=\textwidth]{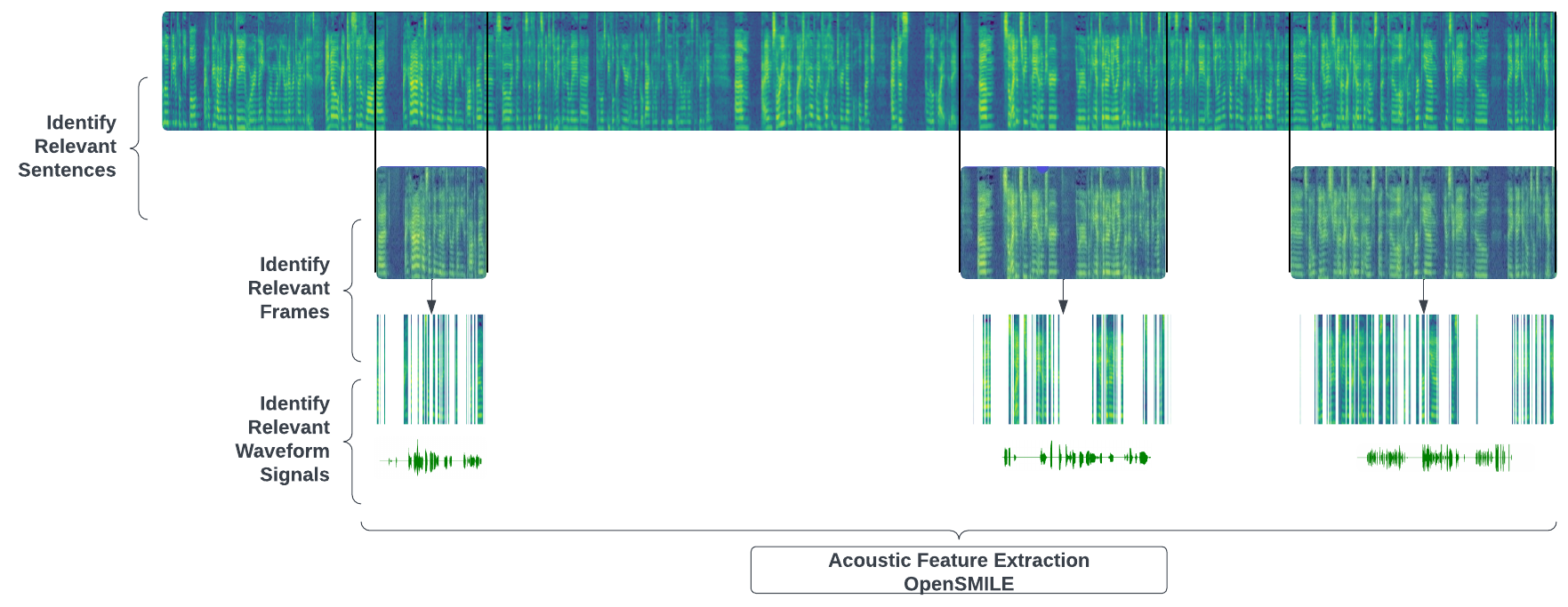}
  \caption{Workflow of the frame-based attention interpretation method. For demonstration purposes, the long spectrogram represents a long speech interval of ten sentences as an input for the proposed model. Here, the speech-level interpretation first identifies the most relevant five sentences, with indexes of 2, 5, 7, 8, and 9. Then, the sentence-level interpretation identifies the relevant frames, using a relevancy threshold of 0.3, for each sentence. Lastly, the waveform signals that temporally correspond to the relevant frames are identified, which are thereby processed by OpenSMILE for relevant acoustic feature extraction.}
  \label{fig:speech_production}
\end{figure*}

\section{Method}
\label{sec:method}

\subsection{Data pre-processing}
The D-vlog dataset \cite{yoon_d-vlog_2022} was used. To obtain audio segments consisting of natural sentences, we first obtained text data by applying the open-sourced Whisper model \cite{radford_robust_2023} to transcribe the waveforms into texts with word-level time stamps. We then designed a sentence-level data segmentation approach to obtain text segments consisting of natural sentences, each with a number of words generally longer than seven. 

Each text segment has word-level timestamps in millisecond units, indicating the start and end times of each sentence. These timestamps are utilised to retrieve the relative sentence-level waveforms to constitute the audio segments, with temporal lengths averaging approximately 5.53s (standard deviation, sd = 13.53s). Each waveform is transformed into a sequence of 128-dimensional log Mel filterbank features, utilizing a 25ms Hamming window with a step size of 10ms. The resulting Mel-spectrograms were uniformly padded or truncated to dimensions of $128 \times 1024$.

\subsection{Speech-level Audio Spectrogram Transformer}


The proposed speech foundation model approach is based entirely on the attention mechanism \cite{vaswani_attention_2017}. First, the model consists of a sentence-level processing block, which encodes sentence-level spectrograms using a pre-trained frame-based AST. Instead of partitioning the spectrogram into patches of size 16 \texttimes{} 16, the frame-based AST splits the spectrogram into frames of size 128 \texttimes{} 2 along the temporal dimension. We add attention masks to ignore padding. After the processing, the \verb|[cls]| token from the audio data then represents the sentence-level audio data.

Second, the model consists of a speech-level processing block which is also a transformer encoder (six consecutive attention layers) that operates at the speech level. It receives the sequence of sentence representations produced from the sentence-level processing block for each participant and injects positional embeddings to consider the sequence order of each sentence. A speech-level \verb|[cls]| token is prepended to the sequence to aggregate the sequence into a single representation which will be mapped onto a 2-dimensional space for the final binary classification.

Consider a given speech $S_i = \{s_{i1}, s_{i2}, ..., s_{in}\}$ from a participant $P_i$, where $s_{ij}$ represents the Mel-spectrogram of the $j^{th}$ sentence in the speech. The speech $S_i$ is passed through the sentence-level processing block to be processed by the sentence-level encoders into a sequence of embeddings $E_i = \{e_{i1}, e_{i2}, ..., e_{in}\}$, representing each sentence in the speech of participant $P_i$. The sequence $E_i$ is then encoded by the speech-level processing block into a single representation $r_i$ for binary classification.

\subsection{Frame-based attention interpretation}

Since our model has two processing blocks (i.e., sentence-level and speech-level), our frame-based interpretation method is hierarchical to first provide a speech-level interpretation, addressing the question ``\textit{Which sentences within a given speech are most relevant to depression detection?}'', and then a sentence-level interpretation, addressing the question: ``\textit{Within the relevant sentences, which Mel-spectrogram frames are most relevant to depression detection?}''. A visualisation of the interpretation procedure can be seen in Figure \ref{fig:speech_production}

For the speech-level interpretation, we derive attention scores from within the speech-level processing block. For each attention layer, we apply the approach introduced in \cite{chefer_generic_2021} to weigh the relative importance of attention scores across attention heads to obtain a gradient-weighted attention map $\bar{A}$, set out in equation (\ref{average attention heads}), where $\nabla A:= \frac{\partial y_d}{\partial A}$ represents the gradients of the output for the depression class $d$ with respect to the attention scores $A$. The Hadamard product $\odot$ accounts for the relative importance of attention scores. We take the mean $\boldsymbol{E}_h$ across heads, with negative contributions removed.

\begin{equation}
\bar{A} = \boldsymbol{E}_h[(\nabla A \odot A)^+]
\label{average attention heads}
\end{equation}

We initialise a relevancy map with the identity matrix for the speech-level processing block as \( R^{ss} = \boldsymbol{I}^{s \times s} \), considering each sentence representation and the speech-level \verb|[cls]| token as initially ``self-relevant''. Then, we apply equation (\ref{self-attention: unimodal}), where \( XX \) can represent either \( ss \) or \( aa \), to update \( R^{ss} \) with a forward pass across every self-attention layer within the block. This provides a mechanism for continuously tracking relevancy between representations at deeper layers while updating the relevancy map.

\begin{equation}
R^{XX} \leftarrow R^{XX} + \bar{A} \cdot R^{XX}
\label{self-attention: unimodal}
\end{equation}

After updating, we take the first row of the matrix $R^{ss}$, corresponding to the position of the \verb|[cls]| token, which contains a relevancy score for each sentence representation. We interpret the sentences at the positions with the highest relevancy scores as most relevant to depression detection. 

We then perform the sentence-level interpretation for the most relevant sentence representations. We first derive the attention scores with respect to these representations from within the sentence-level processing block. Next, we follow the same procedure as above to obtain a gradient-weighted attention map $\bar{A}$ from each attention layer.

We initialise a relevancy map \( R^{aa} = \boldsymbol{I}^{a \times a} \) to account for the self-attention interactions within the audio modality (i.e. between spectrogram frames). We then apply equation (\ref{self-attention: unimodal}) to update $R^{aa}$ across the attention layers within the sentence-level processing block with a forward pass.

After updating, we extract the first row from the relevancy map \( R^{aa} \), corresponding to the position of the sentence-level \verb|[cls]| token. This row \( \mathbf{r} = [r_1, r_2, \ldots, r_n] \) contains a relevancy score for every Mel-spectrogram frame \( n \) within the sentence. We then apply the Min-Max normalization process to this row, adjusting each element to have a value ranging between 0 and 1. The normalised relevancy scores are used to highlight the frames that are relevant to depression detection.

Lastly, we identify the waveform signals corresponding temporally to the relevant spectrogram frames, which are processed using OpenSMILE \cite{eyben_opensmile_2010} to extract relevant acoustic features that are interpretable to human experts.

\begin{figure*}[!t]
  \centering
  \includegraphics[width=\textwidth]{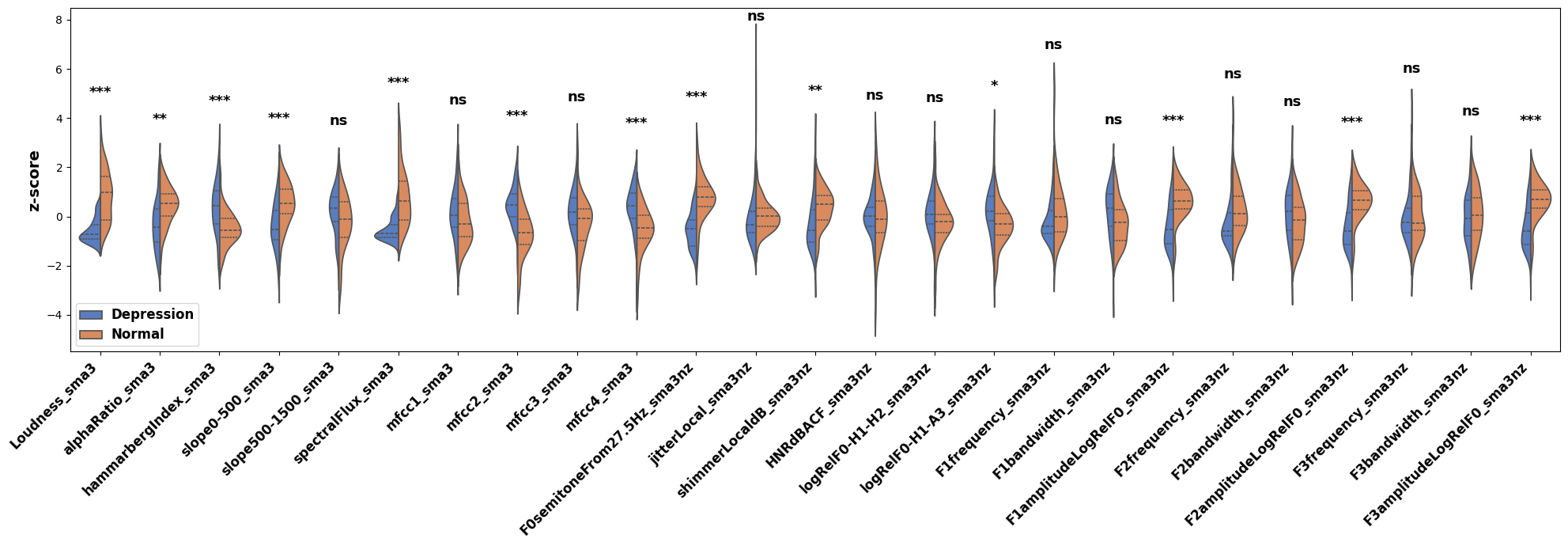}
  \caption{Violin plot of relevant acoustic feature value distributions (residualized for sex and standardized) between true positives (\textit{n = 60}) and true negatives (\textit{n = 41}). Relevant acoustic features were extracted from the waveform signals temporally corresponding to the spectrogram frames with relevancy scores higher than 0.3 from each sample's five most relevant sentences. For true negatives, gradients of the output for the ``normal'' class regarding the attention scores were used to weigh the attention maps. Statistical significance was assessed using the Mann-Whitney U test, with Bonferroni correction applied for multiple comparisons. Significance levels were denoted as: \(ns\) for not significant; * for \(p \leq 0.002\), indicating significant differences; ** for \(p \leq 0.0004\); and *** for \(p \leq 4 \times 10^{-5}\).}

  \label{fig: violin plot}
\end{figure*}

\section{Experiments}
\label{sec:experiments}

\subsection{Data}
\label{data}
D-Vlog, the dataset used in our experiments, consists of 961 YouTube video vlogs labelled by trained annotators as either ``depressed'' or ``normal''. The authors shared the YouTube video keys which were used to download the audio waveforms for our research purposes. However, some videos were made unavailable by their owners. Also, due to limited computational resources, we did not consider videos that are longer than 15 minutes. In total, we downloaded 698 waveforms, with 52.7\% labelled as ``depressed''. We stratified data splitting based on unique YouTube accounts to prevent data leaks. We also stratified based on class and gender to produce the train/development splits shown in Table \ref{tab: sets distribution}, using an 8:2 ratio. One ``depressed'' sample with data processing issues was excluded from the development set.

\begin{table}[h]
  \caption{Number of Depression and Normal instances per set}
  \label{tab: sets distribution}
  \centering
  \begin{tabular}{lcc}
    \toprule
    \textbf{}               & \textbf{Depression} & \textbf{Normal} \\
    \midrule
    Train                   & 294                 & 261            \\
    Development                    & 73                  & 69             \\
    \bottomrule
  \end{tabular}
\end{table}

\subsection{Segment-level model}

To investigate if segment-level labelling noise affects audio modality for depression detection, we fine-tuned a frame-based AST to make segment-level predictions (i.e. lacking the speech-level processing block compared to the proposed speech-level model). We used the first 42 sentences segmented from each participant's speech, each labelled as either ``depressed'' or ``normal'' according to the participant's overall label. Note that some participants have fewer than 42 sentences in their speech (mean = 39.41, sd = 11.42). To derive a speech-level probabilistic prediction for each participant, we calculate the predicted probability of depression as the ratio of sentences classified as ``depressed" to the total number of speech sentences for that participant

\subsection{Implementation details}

One NVIDIA A100 GPU was used for conducting experiments. We trained the proposed speech-level model also using the first 42 sentences segmented from each participant's speech. The sentence-level spectrograms from each participant were batched for the sentence-level processing block to process, which outputs one sequence of embeddings, with a minibatch size of 1, for the speech-level processing block to process. We applied gradient accumulation, whereby gradients are accumulated for 72 training steps before each parameter update. 

For both the speech-level and segment-level models, we froze the first nine layers and randomly reinitialised the last three layers of the pre-trained AST. Both models were trained using an Adam optimizer \cite{kingma_adam_2017} with a learning rate of $3 \times 10^{-5}$ and a weight decay of 0.01. For a fair comparison, training was stopped at the onset of overfitting (epoch 9 for the speech-level model and epoch 8 for the segment-level model) as indicated by the cessation of further decrease in development loss.\footnote{The maximum number of epochs was set to 10 for the speech-level model and 15 for the segment-level model. As the learning rate schedule implements a gradual decay over the total number of epochs, this results in slightly different learning rate schedules between the two models.} While we acknowledge that this approach may lead to overly optimistic estimates of performance for both models (lacking an independent test set due to limited sample size), we note that this work does \textit{not} aim to benchmark performance on the D-Vlog dataset. Instead, we aim to provide robust evidence for the advantage of using a patient’s long speech over short segments for more reliable depression detection in clinical practice.

\subsection{Model performance}

Table~\ref{tab:auc_performance} displays the AUC (Area Under the Receiver Operating Characteristics Curve) scores for the segment-level and speech-level models evaluated on the development set. The speech-level model demonstrates a statistically significant improvement in AUC score over the segment-level model ($p < 0.05$). The result suggests the presence of segment-level labelling noise in audio modality and the advantage of longer-duration speech analysis, which avoids using segment-level labelling, for more reliable depression detection in clinical practice.

\begin{table}[h]
\centering
\caption{AUC performances for models, including DeLong's test p-value for statistical comparison. AUC serves as a classification threshold-invariant measure of model discrimination ability (see a detailed explanation of AUC in \cite{huang2005using}). The 95\% confidence intervals (CI) for AUC scores were obtained through bootstrapping the development set (5,000 bootstrap samples), using the ConfidenceIntervals package (v0.0.3) \cite{FerrerRiera2023}.}
\begin{tabular}{lccc}
\hline
\textbf{Model} & \textbf{AUC (95\% CI)} & \textbf{p-value} \\
\hline
Segment-level & 0.714 [$-.085, +.082$] & \textemdash \\
Speech-level & 0.772 [$-.080, +.074$] & 0.0127 \\
\hline
\end{tabular}
\label{tab:auc_performance}
\end{table}

\begin{figure}[!t]
  \centering
  \includegraphics[width=\linewidth]{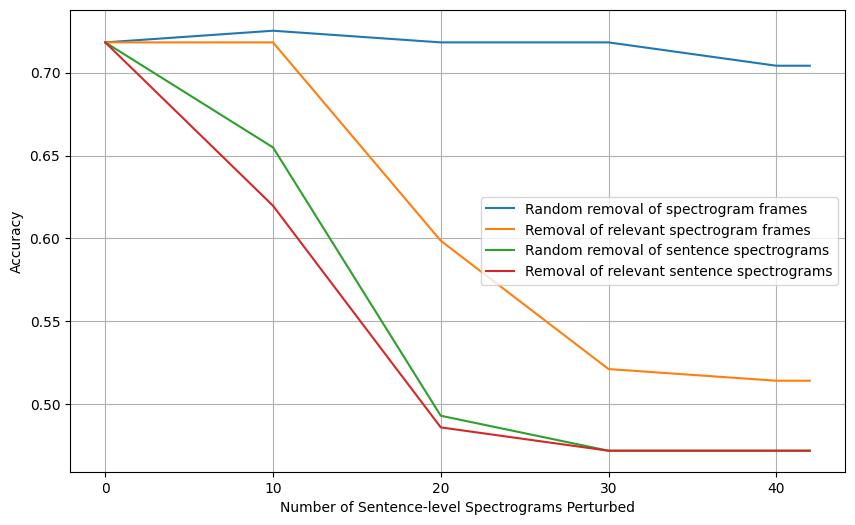}
  \caption{Perturbation test results for the proposed model. Accuracy was computed using a decision threshold of 0.527, reflecting the depression prevalence in the data. In the first test, sentence-level spectrograms were incrementally excluded in descending order of relevance, 10 at a time. In the second, only frames with a relevance score above 0.3 within these sentences were removed. The results of these two tests were benchmarked against random exclusions: one with random sentence spectrogram exclusions, and another with random frames exclusions (30\% within relevant sentence spectrogrames). Accuracy never drops below 47.18\%, likely due to the model’s bias towards positive predictions in the presence of a slight class imbalance.}
  \label{permutation results}
\end{figure}

\subsection{Perturbation study}

We conducted perturbation tests to explore the impact of omitting relevant sentence-level spectrograms or frames on our speech-level model's performance. As shown in Figure \ref{permutation results}, with detailed test descriptions provided in the figure caption, the selective removal of relevant frames led to a marked decline in model accuracy, substantially outpacing the accuracy degradation observed with random frame removal. The selective exclusion of relevant sentence-level spectrograms resulted in a further reduction in accuracy, which is marginally more pronounced, especially for the first 10 sentences, than that observed with random removal of sentences. This observation implies differences between sentence-level spectrograms, or speech segments, in terms of signal importance for depression detection.

\subsection{Relevant acoustic feature extraction}

For the final step of the interpretation method, we locate the waveform signals temporally corresponding to the relevant spectrogram frames and extract a set of acoustics features from these signals using OpenSMILE (v2.5.0). We used the eGeMAPS (v02) feature set, which was designed to index affective changes in voice \cite{eyben_geneva_2016}. Figure \ref{fig: violin plot} demonstrates the distribution differences in relevant acoustic feature values between true positive/negative samples. For a preliminary interpretation, we note that the model identifies reductions in loudness and fundamental frequency (F0) as relevant signals of depression, which aligns with the speech characteristics of depressed patients documented in clinical studies \cite{cummins_review_2015}. Future work is planned to validate the clinical relevance of our method using clinical speech datasets of depression (e.g., DAIC-WOZ \cite{gratch2014distress}).

\section{Conclusions}

This work presents an interpretable speech foundation model approach that leverages pre-trained parameters, enabling scalability with more data, and uses patients' long speech for more reliable depression detection in clinical practice. Crucially, it advances responsible AI by introducing a feasible audio interpretation method that reveals prediction-relevant acoustic features for clinician interpretation, which enhances early screening tools' clinical applicability. With such interpretability, future works may systematically study the behaviour of speech-based depression detection systems. For instance, a system's characteristics (e.g., reliance on certain acoustic features) in making false positives/negatives can be documented prior to their clinical implementations, which allows clinicians to critically assess the reliability of certain model predictions.

\section{Acknowledgements}
This work was supported by UKRI Grant No. 10102226 (University of
Edinburgh) as part of the Horizon Europe Guarantee funding for
participation in the INT-ACT project, funded by REA under the Horizon
Europe Programme, grant 101132719.

\bibliographystyle{IEEEtran}
\bibliography{mybib}

@misc{FerrerRiera2023,
  author = {Ferrer, L. and Riera, P.},
  title = {Confidence Intervals for Evaluation in Machine Learning},
  howpublished = {\url{https://github.com/luferrer/ConfidenceIntervals}},
  note = {[Computer software]},
  year = {2023}
}

@article{huang2005using,
  title={Using {AUC} and accuracy in evaluating learning algorithms},
  author={Huang, Jin and Ling, Charles X},
  journal={IEEE Transactions on knowledge and Data Engineering},
  volume={17},
  number={3},
  pages={299--310},
  year={2005},
  publisher={IEEE}
}

@article{eyben_geneva_2016,
	title = {The {Geneva} {Minimalistic} {Acoustic} {Parameter} {Set} ({GeMAPS}) for {Voice} {Research} and {Affective} {Computing}},
	volume = {7},
	issn = {1949-3045},
	url = {https://ieeexplore.ieee.org/abstract/document/7160715},
	doi = {10.1109/TAFFC.2015.2457417},
	number = {2},
	urldate = {2024-02-23},
	journal = {IEEE Transactions on Affective Computing},
	author = {Eyben, Florian and Scherer, Klaus R. and Schuller, Björn W. and Sundberg, Johan and André, Elisabeth and Busso, Carlos and Devillers, Laurence Y. and Epps, Julien and Laukka, Petri and Narayanan, Shrikanth S. and Truong, Khiet P.},
	month = apr,
	year = {2016},
	note = {Conference Name: IEEE Transactions on Affective Computing},
	pages = {190--202},
}

@inproceedings{eyben_opensmile_2010,
	address = {New York, NY, USA},
	series = {{MM} '10},
	title = {Opensmile: the munich versatile and fast open-source audio feature extractor},
	isbn = {978-1-60558-933-6},
	shorttitle = {Opensmile},
	url = {https://dl.acm.org/doi/10.1145/1873951.1874246},
	doi = {10.1145/1873951.1874246},
	urldate = {2023-12-26},
	booktitle = {Proceedings of the 18th {ACM} international conference on {Multimedia}},
	publisher = {Association for Computing Machinery},
	author = {Eyben, Florian and Wöllmer, Martin and Schuller, Björn},
	month = oct,
	year = {2010},
	pages = {1459--1462},
}

@online{WHO2022,
    author = {{World Health Organization}},
    title = {Depressive disorder (depression)},
    year = {2023},
    howpublished = {https://www.who.int/news-room/fact-sheets/detail/depression},
    note = {Accessed: August 15, 2023}
}

@article{gong_ssast_2022,
	title = {{SSAST}: {Self}-{Supervised} {Audio} {Spectrogram} {Transformer}},
	volume = {36},
	copyright = {Copyright (c) 2022 Association for the Advancement of Artificial Intelligence},
	issn = {2374-3468},
	shorttitle = {{SSAST}},
	url = {https://ojs.aaai.org/index.php/AAAI/article/view/21315},
	doi = {10.1609/aaai.v36i10.21315},
	language = {en},
	number = {10},
	urldate = {2023-12-23},
	journal = {Proceedings of the AAAI Conference on Artificial Intelligence},
	author = {Gong, Yuan and Lai, Cheng-I. and Chung, Yu-An and Glass, James},
	month = jun,
	year = {2022},
	note = {Number: 10},
	pages = {10699--10709},
}

@article{yoon_d-vlog_2022,
	title = {D-vlog: {Multimodal} {Vlog} {Dataset} for {Depression} {Detection}},
	volume = {36},
	copyright = {Copyright (c) 2022 Association for the Advancement of Artificial Intelligence},
	issn = {2374-3468},
	shorttitle = {D-vlog},
	url = {https://ojs.aaai.org/index.php/AAAI/article/view/21483},
	doi = {10.1609/aaai.v36i11.21483},
	language = {en},
	number = {11},
	urldate = {2023-07-14},
	journal = {Proceedings of the AAAI Conference on Artificial Intelligence},
	author = {Yoon, Jeewoo and Kang, Chaewon and Kim, Seungbae and Han, Jinyoung},
	month = jun,
	year = {2022},
	note = {Number: 11},
	pages = {12226--12234},
}

@article{sardari_audio_2022,
	title = {Audio based depression detection using {Convolutional} {Autoencoder}},
	volume = {189},
	issn = {0957-4174},
	url = {https://www.sciencedirect.com/science/article/pii/S0957417421014147},
	doi = {10.1016/j.eswa.2021.116076},
	language = {en},
	urldate = {2023-07-14},
	journal = {Expert Systems with Applications},
	author = {Sardari, Sara and Nakisa, Bahareh and Rastgoo, Mohammed Naim and Eklund, Peter},
	month = mar,
	year = {2022},
	pages = {116076},
}

@inproceedings{vaswani_attention_2017,
	title = {Attention is All you Need},
	volume = {30},
	url = {https://proceedings.neurips.cc/paper_files/paper/2017/hash/3f5ee243547dee91fbd053c1c4a845aa-Abstract.html},
	urldate = {2023-07-14},
	booktitle = {Advances in {Neural} {Information} {Processing} {Systems}},
	publisher = {Curran Associates, Inc.},
	author = {Vaswani, Ashish and Shazeer, Noam and Parmar, Niki and Uszkoreit, Jakob and Jones, Llion and Gomez, Aidan N and Kaiser, Lukasz and Polosukhin, Illia},
	year = {2017},
}

@article{squires_deep_2023,
	title = {Deep learning and machine learning in psychiatry: a survey of current progress in depression detection, diagnosis and treatment},
	volume = {10},
	issn = {2198-4026},
	shorttitle = {Deep learning and machine learning in psychiatry},
	url = {https://doi.org/10.1186/s40708-023-00188-6},
	doi = {10.1186/s40708-023-00188-6},
	number = {1},
	urldate = {2023-07-14},
	journal = {Brain Informatics},
	author = {Squires, Matthew and Tao, Xiaohui and Elangovan, Soman and Gururajan, Raj and Zhou, Xujuan and Acharya, U Rajendra and Li, Yuefeng},
	month = apr,
	year = {2023},
	pages = {10},
}

@article{muzammel_end--end_2021,
	title = {End-to-end multimodal clinical depression recognition using deep neural networks: {A} comparative analysis},
	volume = {211},
	issn = {0169-2607},
	shorttitle = {End-to-end multimodal clinical depression recognition using deep neural networks},
	url = {https://www.sciencedirect.com/science/article/pii/S0169260721005071},
	doi = {10.1016/j.cmpb.2021.106433},
	language = {en},
	urldate = {2023-07-14},
	journal = {Computer Methods and Programs in Biomedicine},
	author = {Muzammel, Muhammad and Salam, Hanan and Othmani, Alice},
	month = nov,
	year = {2021},
	pages = {106433},
}

@article{rudin_stop_2019,
	title = {Stop explaining black box machine learning models for high stakes decisions and use interpretable models instead},
	volume = {1},
	copyright = {2019 Springer Nature Limited},
	issn = {2522-5839},
	url = {https://www.nature.com/articles/s42256-019-0048-x},
	doi = {10.1038/s42256-019-0048-x},
	language = {en},
	number = {5},
	urldate = {2023-07-14},
	journal = {Nature Machine Intelligence},
	author = {Rudin, Cynthia},
	month = may,
	year = {2019},
	note = {Number: 5
Publisher: Nature Publishing Group},
	pages = {206--215},
}

@misc{bucur_end--end_2022,
	title = {An {End}-to-{End} {Set} {Transformer} for {User}-{Level} {Classification} of {Depression} and {Gambling} {Disorder}},
	url = {http://arxiv.org/abs/2207.00753},
	doi = {10.48550/arXiv.2207.00753},
	urldate = {2023-07-14},
	publisher = {arXiv},
	author = {Bucur, Ana-Maria and Cosma, Adrian and Dinu, Liviu P. and Rosso, Paolo},
	month = jul,
	year = {2022},
	note = {arXiv:2207.00753 [cs]},
}

@article{zogan_explainable_2022,
	title = {Explainable depression detection with multi-aspect features using a hybrid deep learning model on social media},
	volume = {25},
	issn = {1573-1413},
	url = {https://doi.org/10.1007/s11280-021-00992-2},
	doi = {10.1007/s11280-021-00992-2},
	language = {en},
	number = {1},
	urldate = {2023-07-14},
	journal = {World Wide Web},
	author = {Zogan, Hamad and Razzak, Imran and Wang, Xianzhi and Jameel, Shoaib and Xu, Guandong},
	month = jan,
	year = {2022},
	pages = {281--304},
}

@misc{jain_attention_2019,
	title = {Attention is not {Explanation}},
	url = {http://arxiv.org/abs/1902.10186},
	doi = {10.48550/arXiv.1902.10186},
	urldate = {2023-07-14},
	publisher = {arXiv},
	author = {Jain, Sarthak and Wallace, Byron C.},
	month = may,
	year = {2019},
	note = {arXiv:1902.10186 [cs]},
}

@misc{wiegreffe_attention_2019,
	title = {Attention is not not {Explanation}},
	url = {http://arxiv.org/abs/1908.04626},
	doi = {10.48550/arXiv.1908.04626},
	urldate = {2023-07-14},
	publisher = {arXiv},
	author = {Wiegreffe, Sarah and Pinter, Yuval},
	month = sep,
	year = {2019},
	note = {arXiv:1908.04626 [cs]},
}

@inproceedings{chefer_generic_2021,
	title = {Generic {Attention}-{Model} {Explainability} for {Interpreting} {Bi}-{Modal} and {Encoder}-{Decoder} {Transformers}},
	url = {https://openaccess.thecvf.com/content/ICCV2021/html/Chefer_Generic_Attention-Model_Explainability_for_Interpreting_Bi-Modal_and_Encoder-Decoder_Transformers_ICCV_2021_paper.html},
	language = {en},
	urldate = {2023-07-14},
	author = {Chefer, Hila and Gur, Shir and Wolf, Lior},
	year = {2021},
	pages = {397--406},
}

@inproceedings{radford_robust_2023,
	title = {Robust {Speech} {Recognition} via {Large}-{Scale} {Weak} {Supervision}},
	url = {https://proceedings.mlr.press/v202/radford23a.html},
	language = {en},
	urldate = {2023-07-14},
	booktitle = {Proceedings of the 40th {International} {Conference} on {Machine} {Learning}},
	publisher = {PMLR},
	author = {Radford, Alec and Kim, Jong Wook and Xu, Tao and Brockman, Greg and Mcleavey, Christine and Sutskever, Ilya},
	month = jul,
	year = {2023},
	note = {ISSN: 2640-3498},
	pages = {28492--28518},
}

@misc{kingma_adam_2017,
	title = {Adam: A Method for Stochastic Optimization},
	shorttitle = {Adam},
	url = {http://arxiv.org/abs/1412.6980},
	doi = {10.48550/arXiv.1412.6980},
	urldate = {2023-07-18},
	publisher = {arXiv},
	author = {Kingma, Diederik P. and Ba, Jimmy},
	month = jan,
	year = {2017},
	note = {arXiv:1412.6980 [cs]},
}

@article{cummins_review_2015,
	title = {A review of depression and suicide risk assessment using speech analysis},
	volume = {71},
	issn = {0167-6393},
	url = {https://www.sciencedirect.com/science/article/pii/S0167639315000369},
	doi = {10.1016/j.specom.2015.03.004},
	language = {en},
	urldate = {2023-08-07},
	journal = {Speech Communication},
	author = {Cummins, Nicholas and Scherer, Stefan and Krajewski, Jarek and Schnieder, Sebastian and Epps, Julien and Quatieri, Thomas F.},
	month = jul,
	year = {2015},
	pages = {10--49},
}

@inproceedings{gratch2014distress,
  title={The distress analysis interview corpus of human and computer interviews.},
  author={Gratch, Jonathan and Artstein, Ron and Lucas, Gale M and Stratou, Giota and Scherer, Stefan and Nazarian, Angela and Wood, Rachel and Boberg, Jill and DeVault, David and Marsella, Stacy and others},
  booktitle={LREC},
  pages={3123--3128},
  year={2014},
  organization={Reykjavik}
}

\end{document}